\pdfoutput=1
\documentclass[
  a4paper,
  top=3cm,
  bottom=2cm,
  left=3cm,
  right=3cm,
  marginparwidth=1.75cm
]{article}

\usepackage[letterpaper,margin=20mm]{geometry}
\usepackage[T1]{fontenc}
\usepackage[english]{babel}

\usepackage{newtxtext,newtxmath}

\usepackage{graphicx}
\usepackage{booktabs}
\usepackage{multirow}
\usepackage{array,tabularx}
\usepackage{float}

\usepackage[style=apa,backend=bibtex,defernumbers=false]{biblatex}

\usepackage[colorlinks=true,allcolors=blue]{hyperref}

\newcommand{\footremember}[2]{%
  \footnote{#2}%
  \newcounter{#1}\setcounter{#1}{\value{footnote}}%
}
\newcommand{\footrecall}[1]{\footnotemark[\value{#1}]}

\title{\LARGE Exploring temporal dynamics in digital trace data: mining
user-sequences for communication research\thanks{Preprint version. Manuscript is currently under peer review.}}

\author{%
 Yangliu Fan\footremember{wi}{Weizenbaum Institute for Networked Society, Berlin, Germany} \footremember{fu}{Institute for Media and Communication Studies, Free University of Berlin, Berlin, Germany} \footnote{Corresponding author: Yangliu Fan, yangliu.fan@weizenbaum-institut.de}%
  \and Jakob Ohme\footrecall{wi}%
  \and Lion Wedel\footrecall{wi}}%

\date{}
\begin{document}
\maketitle

\begin{abstract}
Communication is commonly considered a process that is dynamically
situated in a temporal context. However, there remains a disconnection
between such theoretical dynamicality and the non-dynamical character of
communication scholars\textquotesingle{} preferred methodologies. In
this paper, we argue for a new research framework that uses
computational approaches to leverage the fine-grained timestamps
recorded in digital trace data. In particular, we propose to maintain
the hyper-longitudinal information in the trace data and analyze
time-evolving \textquotesingle user-sequences,\textquotesingle{} which
provide rich information about user activity with high temporal
resolution. To illustrate our proposed framework, we present a case
study that applied six approaches (e.g., sequence analysis, process
mining, and language-based models) to real-world user-sequences
containing 1,262,775 timestamped traces from 309 unique users, gathered
via data donations. Overall, our study suggests a conceptual
reorientation towards a better understanding of the temporal dimension
in communication processes, resting on the exploding supply of digital
trace data and the technical advances in analytical approaches.
\end{abstract}

\hspace{10pt}

\section{Introduction}
In today\textquotesingle s fast-paced media environment,
users\textquotesingle{} digital routines generate sequences of
continuous interactions across multiple channels, platforms, and
devices---leaving behind \emph{digital traces}. These ever-generating
trace data range from browser cookies, sensor readings, search queries,
and app logs, to the vast expanse of textual, audio, and visual content
on social media (\cite{Ohme2024, Stier2020a}). As the scope
and resolution of data accessible for research have expanded
dramatically, this presents exciting research opportunities to
communication scholars. However, it also poses serious operational and
technical challenges---that is, once we obtain thousands of digital
points \emph{per user} over an extended period of time on a
second-by-second basis, how do we leverage the potential of this
hyper-longitudinal data at the user level (cf., the experience sampling
method (ESM) and self-reported data) to gain new insights into
communication phenomena?

In this paper, we concur with the observation by \cite{Wells2019} (p.
4022) that "we are standing on the cusp of a new phase in the
field\textquotesingle s opportunities to incorporate temporal dynamics
into communication research." In particular, we propose a new research
framework that employs the longitudinal information recorded in digital
trace data (i.e., "the fine-grained timestamps in the data" (\cite{Lee2023}) to systematically study the evolving and dynamic nature of
user interactions with the media environment. Specifically, our
framework represents the evolving user activities as
\textquotesingle user-sequences\textquotesingle{} with high temporal
resolution. This framework builds on two recent advances: \emph{digital
trace data (DTD)} and \emph{computational approaches.} We believe that
as DTD captures fine-grained, precise, and continuous information people
encounter, when combined with increasingly accessible computational
approaches, it accordingly opens new possibilities for advancing the
conceptualization of \emph{time} in theory building and empirical
investigations in communication research.

Currently, despite the growing research interest in DTD (e.g., \cite{Jungherr2017a,Stier2020a,Theocharis2021,Ohme2024, Otto2024,wu2021}), it has been found that the
longitudinal aspect is often overlooked (\cite{Lee2023}). Instead,
this detailed temporal information is typically aggregated into
frequency counts over a specific time period, e.g., daily online news
exposure (\cite{Cronin02102023}). Or, researchers collapse a series of
event logs to concentrate on the interaction results as a static
pair-wise relationship between entities (\cite{Keegan2016}). This is
because most existing measures in the field are rooted in the so-called
\emph{time-budget paradigm,} when temporal information is limited  (\cite{Lee2023}). Yet, DTD provides rich temporal information about user
activity, allowing the application of many new methods for capturing
temporal dynamics. Few pioneering studies (e.g., \cite{Yeykelis2014,Wells2019,Hopp2020,Vermeer2020,Reeves2021,Wu2021a}) have begun to transcend the time-budget
paradigm by drawing methodological expertise and inspiration from other
fields. For example, a recent study uses Hidden Markov Models (HMM) to
explore the sequential transitions between news records (\cite{Hopp2020}). Although many approaches exist for analyzing DTD, the method
research on the \emph{second-order} behavioral attributes (e.g.,
temporal patterns, path dependencies, and behavioral motifs) remains
limited, compared to the first-order attributes such as activity counts,
content types, and basic user statistics (\cite{Keegan2016}). To date,
much remains to be understood about the analytical possibilities of DTD,
as well as how new approaches might deepen our understanding of complex
communication processes.

Motivated by these gaps in our knowledge, this paper sets out to explore
how to extract meaningful patterns from high-density digital trace data
while preserving temporal details (without considerable temporal
aggregation). This has constantly been a challenge in communication
research, where the aggregation of temporal dynamics into
cross-sectional measures is more common (\cite{Baumgartner2024}). Hence, this
study proposes a framework that leverages the temporal resolution in the
trace data and studies time-evolving
\textquotesingle user-sequences,\textquotesingle{} which allows us to
discover moment-by-moment transitions and how they evolve into long-term
patterns (\cite{Reeves2021}). We consider our framework to provide a
flexible lens that allows us to zoom in and out across analytical units
and time scales (e.g., individual-level processes vs. group-level
behaviors, individual switches vs. whole sequences). Moreover, this
paper also discusses important analytical challenges brought by
high-density trace data. By focusing on an original dataset of over a
million digital traces across four social media platforms, we show that
with hyper-longitudinal data, a highly detailed picture of
individual-level patterns emerges, and different approaches (e.g.,
sequence analysis, process mining, and language-based models) uncover
this picture at multiple levels of granularity. Practically, this study
offers an overview and initial guidance for communication scholars new
to exploring the second-order attributes in the DTD. In summary, we
believe that "user-sequences" as a new framework has the potential to
advance our field through a conceptual reorientation of the \emph{time}
dimension in communication research (\cite{Wells2019}).

\section{The theoretical framework and related research}
\label{sec:headings}
\subsection{\emph{\textbf{Rethink time in communication research}}}

Time---a multidimensional object that includes aspects such as timing,
sequence, and interval---is considered the universal factor underlying
all human behaviors (\cite{Flaherty2003,Lee2023}). Following the
axiom that communication is a \emph{process}, the major theories and
models in our field consider time a central dimension of communication.
However, there has been a longstanding gap between the axiomatic idea of
communication as a process and the actual exploitation of the time
dimension in communication research (\cite{Yanovitzky2008,Wells2019,Brinberg2023}).

Much of the existing research has overlooked the function of time as
they aggregated temporal information in the data (as outlined by \cite{Lee2023}). Such temporal simplification inherently assumes
\emph{temporal stability} in the data. This assumption, however, is
limited, as large amounts of data generated during brief, intense
moments may overshadow subtle patterns during long, non-intense periods
(\cite{Wells2019}). Moreover, it fails to account for characteristics
of time as a refined concept, including aspects such as timing,
sequence, and intervals (\cite{Flaherty2003}). Additionally, it is thought
that maintaining temporal details allows scholars to reliably examine
the intraindividual processes---within-person changes over time
(\cite{Brinberg2023,Otto2024}). This moves beyond
the traditional cross-sectional designs, which are considered to be
relatively limited when establishing \emph{causal} relationships.

Even so, it is believed that many major models in our field are causal
in nature as they explicitly focus on the \emph{effects} (\cite{Wells2019}). To establish causality, it is essential to determine the temporal
order of events, i.e., the change in \emph{X} must precede the change
in \emph{Y} in time (\cite{Miller2000}). Despite the significant attention on media
effects and related studies within our field (see \cite{Neuman2011} for a study on the evolution of media effects theory), one
commonly criticized temporal simplification is that most existing effect
metrics assume \emph{linearity} (\cite{Singer2018,Wells2019,Baumgartner2024}).

In response, researchers (\cite{Shehata2021,Otto2024}) have
advocated a more nuanced paradigm of media effects, including
intrapersonal, reciprocal, immediate, and long-term effects. Further,
some effects can be characterized by complex dynamics, such as
stabilization where individual responses to media stimuli stabilize
after repeated exposure and delays where the effects may not become
evident immediately (e.g., sleeper effect) (\cite{Wells2019,Brinberg2023,Baumgartner2024}). In this regard, a recent work
by \cite{Brinberg2023} offers a conceptual overview of
common \emph{(within-person)} \emph{change} processes in communication
phenomena, including order relations and feedback loops. Their study
also briefly describes several analytical approaches---particularly
autoregressive and other regression-based methods---suited to observe
those changes. However, countering these enthusiastic proclamations
about capturing changes on the \emph{intraindividual} level (\cite{Valkenburg2013,Brinberg2023}), a recent study
(\cite{Baumgartner2024}) argues that media effects should instead be
conceptualized as \emph{drifting tectonic plates}, where within-person
changes are mostly undetectable due to the stabilization of the effects
over time.

Indicative as these conceptual studies may be, it is hard to draw a
conclusion on their basis regarding the function of time, either in
principle or for empirical work. As we continue refining our theoretical
frameworks and empirical techniques, it is thus crucial to interrogate
the temporal dimension in communication research. Currently, our field
is only starting to come to terms with the temporal complexity both
theoretically and empirically (\cite{Wells2019,Lee2023}).
This paper presents new opportunities to advance our understanding of
time using digital trace data. It specifically proposes a new research
framework\emph{,} which helps enrich our understanding of how temporal
dynamics play out in complex communication processes.
\subsection{\emph{\textbf{The proposed framework}}}

In this study, we propose a new research framework enabled by the
exploding supply of digital trace data (DTD), alongside various
approaches to analyzing them. We define our framework as an analytical
scheme that \emph{chronologically} orders user interactions with the
media environment into one dimension and employs computational
approaches to explore the inter-activity dependence that considers
individual activities in the \emph{contexts} of other activities.
Specifically, it represents the evolving user activities on digital
platforms as user-sequences, which provide detailed information about
user activity with high temporal resolution (see Figure 1 for an example
of the proposed individual-level representation).

Our framework builds on the idea originating from sequence analysis
(\cite{Hannan1979,Abbott1995,Keegan2016,Mahringer2021,Savcisens2024})---that is, temporal ordering in
the sequences is important, and elements within the sequence have
attributes at multiple levels (e.g., timestamp, platform, and activity
type). In communication science literature, there is growing interest in
sequential and contextual effects (\cite{Vermeer2020,Reeves2021,Wu2021a}). For instance, media use has been
conceptualized as a \emph{flow} in which people encounter a sequence of
offerings (\cite{Wu2021a}), and interpretation and perception of
information may be influenced by what precedes and follows through
processes like priming (\cite{Reeves2021}). Moreover, \cite{Friemel2023} argue that communication actions by actors are embedded
in the context of their preceding actions, where the production or
reception of the current content is dependent on the content that has
been produced or perceived before. Uniquely relevant to the existing
studies on serial switching, time segments, and behavior sequences using
DTD (\cite{Yeykelis2014,Reeves2021}), our framework aims to
provide a meaningful way to capture the holistic picture of
users\textquotesingle{} digital experiences using DTD, including when
people \emph{switch} between contents and how their digital behavior
\emph{evolve} over an extended period of time. Moreover, the
individual-level sequences can be aggregated and clustered to study
\emph{grouped-level} phenomena over longer timespans and diverse
populations. Overall, this framework provides a more flexible lens that
enables both a microscopic, individual-level view and a macroscopic view
of analyzing months or years of trace data across hundreds or thousands
of diverse users.

Having provided the definition of our framework, we now focus on its two
main contributing factors: the exploding supply of DTD and increasingly
accessible computational approaches. DTD is created passively and
collected unobtrusively, providing fine-grained, precise, and continuous
details about behavioral patterns (\cite{Otto2024}), unlike
conventional datasets collected by communication researchers, e.g.,
self-reported survey data. Meanwhile, the rise of new data collection
methods, such as data donation, Application Programming Interfaces
(APIs), screen tracking, and web scraping, has greatly facilitated the
availability of DTD (\cite{Reeves2021,Ohme2024}). This
provides communication scholars with unprecedented access that reaches
deeper into today\textquotesingle s personalized, fragmented, and
dynamic media environment (\cite{Otto2024}).

For instance, the data donation approach requests participants to donate
their digital traces to researchers. It is a user-centric approach that
leverages individuals\textquotesingle{} rights to access and request
their personal data from data controllers (\cite{Ohme2024}).
Therefore, it ensures that researchers obtain informed consent from the
participants and allows them to collect non-public data, such as private
and sensitive data (\cite{Ohme2024,Otto2024}). In the
present study, we illustrate the proposed framework based on a
large-scale dataset of digital traces collected through data download
package donation (See Data section for details).

Another contributing feature of the new framework is the use of
computational approaches to analyze the \emph{temporal}
dimension---i.e., the fine-grained timestamps in the DTD that record
when users engage in specific activities on digital platforms. Nowadays,
rapid advances in computer science and engineering offer a variety of
exciting techniques to analyze these large, complex datasets of digital
traces. Many tools and techniques have proven powerful for encoding
longitudinal data and extracting temporal patterns. However, they remain
largely disconnected from DTD-based communication research---despite the
nascent efforts to synchronize them (e.g., \cite{Hopp2020}). In this
study, we specifically introduce a diverse set of analytical approaches
suited to studying user-sequences for communication research, also
incorporating technical advances from recent years.

\subsection{\emph{\textbf{Practical challenges}}}

Before turning to the analytical possibilities, it is crucial to
consider important practical challenges posed by DTD. Prior research
(\cite{Keegan2016,Liu2016,Fiorio2021,Knight2017,Yanovitzky2008,Wells2019,Cuzzocrea2021,Savcisens2024}) have pointed out that hyper-longitudinal
data presents a series of analytical challenges for researchers who
strive to account for temporal dynamics. In this section, we outline
some of the key challenges when mining user-sequences from
individual-level digital trace data.

\begin{enumerate}
\def\labelenumi{\arabic{enumi}.}
\item
  \textbf{Volume}: user-sequences contain massive amounts of data points
  (e.g., thousands of observations per user), which presents scalability
  issues for data storage and processing (\cite{Liu2016,Knight2017}).
\item
  \textbf{Complexity}: the elements within sequences are multivariate,
  which include both the activity and contextual information, such as
  the textual data and potentially graphical and video data (through
  data linkage and augmentation) (\cite{Liu2016,Cuzzocrea2021,Wedel2024}). Hence, user-sequences are high dimensional (e.g.,
  a large number of variables) and sparse (e.g., many variables
  represent infrequent or rare events) (\cite{Savcisens2024}). In
  addition, they show high cardinality (e.g., thousands of content
  options available for each element)(\cite{Liu2016}).
\item
  \textbf{Incompleteness and skewness}: individuals are not consistently
  observed due to various operational challenges. Consequently, missing
  and incomplete data, or "tracking undercoverage," are common issues in
  DTD (\cite{Fiorio2021,Bosch03042025}). For instance, sequences
  might experience left- or right-censoring, which misses the beginning
  or end of user activities. Or, researchers fail to obtain data from
  all devices and platforms that individuals utilize (\cite{Bosch03042025}). In addition, the number of observations per individual is not
  normally distributed---few individuals have many observations, while
  many individuals only have a few observations.
\item
  \textbf{Time window}: researchers need to define the appropriate
  analytical units to set up the grain size of time windows, such as the
  whole period or per day (\cite{Knight2017}).
\item
  \textbf{Data alignment}: user-sequences record how users interact with
  multiple platforms of diverse affordances, rendering it challenging to
  align observations from multiple sources in a meaningful and
  comparable way (\cite{Keegan2016,Fiorio2021}). In the next step, the temporal granularity in DTD cannot be directly linked to data collected through other approaches (e.g., cross-section
or panel surveys). In particular, many dependent variables, such as
partisanship, cannot be observed or do not vary at low levels (\cite{Wells2019}).
\item
  \textbf{Data volatility:} the platforms can alter the scope and
  details of data access at any time (as they are not incentivized to
  maintain consistency in DTD-based measurements for research purposes)
  (\cite{Lazer2020,Buehling2024,Pearson2024}).
  Consequently, the data collected might not represent the full spectrum
  of activities on these platforms. For example, only some platforms
  (e.g., TikTok and YouTube) record watching activities in our studied
  sample. Further, a recent study (\cite{Pearson2024}) found that
  TikTok has changed its API without making detailed announcements to
  the research community.
\end{enumerate}
Having recognized these challenges, we will now proceed to review
methods for mining user-sequences and revisit these challenges after
presenting the case study (see Discussion and conclusion). It is
important to note that while we applaud recent work using DTD to predict
user behaviors for practical purposes, our study focuses specifically on
advancing empirical reach and informing theory-building in communication
research. Therefore, our method review---while not exhaustive or focused
solely on technical advances---is particularly attentive to
communication research questions for which these applications could
provide new insights.

\section{Method review}
This section will provide an overview of a diverse range of analytical
approaches. The (inter-)disciplinary efforts to study the temporal
aspects are heterogeneous in the existing method research and empirical
work. In this paper, we focus on six approaches that lent themselves to
capturing sequential patterns. These include (\emph{a}) sequence
analysis, (\emph{b}) event history analysis, (\emph{c}) hidden Markov
models, (\emph{d}) network analysis, (\emph{e}) process mining models,
and (\emph{f}) language-based models. Our review will briefly introduce
each approach and discuss how they can be applied to extract information
from user-sequences.

\subsection{\emph{\textbf{Sequence analysis}}}
Sequence analysis has been widely used to identify, compare, and
visualize patterns in temporally ordered data (\cite{Abbott1995, Mahringer2021}). It has been applied in many neighboring fields to
study the longitudinal aspects of human life. Specifically, it adopts a
holistic approach, which treats sequences as whole units to study
patterns among (sub-)sequences (\cite{Abbott1995,Keegan2016}). This
approach is considered particularly powerful at identifying common
sequences and transition patterns, clustering similar sequences, and
visualizing categorical sequences (\cite{Abbott1995,Fasang2014,Keegan2016}).

In the context of the proposed framework, sequence analysis can be used
to identify typical user pathways on digital platforms, cluster similar
users based on their pathways, and uncover nonrandom and complex
behavior motifs. Subsequently, it can be linked to other measures to
determine the origin of these patterns (dependent variables) or how they
influence outcomes of interest (independent variables)(\cite{Abbott1995}).
One possible way is to first categorize users based on their sequential
patterns and then correlate these categories with outcomes measured in
the surveys. Despite its prevalence and applicability, previous research
(\cite{Mahringer2021,Halpin2014,Dlouhy2015})
noted that traditional sequence analysis is limited as it studies
discrete-time sequences as strings (abstract simplifications without
differences in meaning associated with elements) and can be sensitive to
sequence lengths.

\subsection{\emph{\textbf{Event history analysis}}}

Event history analysis, also known as duration analysis or survival
analysis, focuses on the time until transitioning to one significant
event, such as death (\cite{Hannan1979,Abbott1995,Cook2015}). In particular, this approach employs the \emph{duration}
an individual remains in a specific state and the \emph{timing} of
transition in the sequential data. The central concept is the risk
function or hazard rate, which provides a time-varying description of
how the probabilities of the event occurrence change over time
(\cite{Blossfeld2015}).

This approach is believed to be particularly suited to studying the
timing and duration of states in social and behavioral studies (\cite{Brinberg2023}). For instance, \cite{Xing2018} employed
survival analysis to examine users\textquotesingle{} commitment time in
a learning community on Twitter. They found that the more tweets exposed
to the users in the cognitive and interactive dimensions, the lower
their risk of dropping out.

Within the proposed framework, event history analysis can be applied to
examine the transition to one particular state recorded in
user-sequences such as platform switches, which can be linked with
individual-level (or unit-level) covariates measured through other
approaches. However, this method requires a clear definition of the
event in the data and reduces longitudinal phenomena as (discrete) event
patterns. Moreover, prior research (\cite{Mahringer2021}) has
raised concerns regarding inferring the duration of events using
timestamps from DTD, as they only record the starting time of the
events.

\subsection{\emph{\textbf{Hidden Markov Models}}}

Hidden Markov Models (HMM) are stochastic models focusing on one-step
transitions and internal interdependencies within sequences (\cite{Hannan1979,Rabiner1989,Abbott1995}). In particular, HMMs can
identify hidden states where the observation is a probabilistic function
of the state (\cite{Rabiner1989}).

In communication research, HMMs have been employed to predict the next
state in user behavior (time-dependence in the process) and reveal the
hidden states that produce the observed patterns (reconceptualizing the
process in terms of hidden states) (\cite{Hopp2020,Vermeer2020}). For instance, \cite{Hopp2020} employed HMMs to study
the dynamic transactions between events and news frames. They found that
the identified hidden states could encode nuanced relationships between
moral frames and events, which can be used to predict future
occurrences.

Within the proposed framework, HMMs can be employed to understand
underlying mechanisms in user-sequences and predict future states.
However, one important limitation is that Markovian models typically
assume that the probability of being in a given state depends only on
the previous state (but the dependencies can extend through several
states) (\cite{Rabiner1989}). Moreover, they are computationally intensive
(e.g., Baum-Welch algorithms).

\subsection{\emph{\textbf{Network analysis}}}

Network analysis has a long, interdisciplinary history and supports
empirical investigations of interactions among entities both
qualitatively and quantitatively (\cite{Granovetter1973,Scott1988,Hidalgo2016,Decuypere2020}). In communication science, it has been
employed to examine many important topics, including political
polarization (\cite{Yarchi2020}), campaign dynamics (\cite{Nuernbergk2016}), and social movements (\cite{Suitner2023}).

Within the proposed framework, we could construct networks of
sequentially appearing elements from user-sequences. Like other networks
constructed based on communication traces, this approach is flexible and
can be easily extended to many other forms. For example, we could adopt
a dynamic perspective by examining an evolving network of activities at
multiple critical time points. Further, we could examine user-level
patterns based on a certain degree of similarity in co-participating
activities on digital platforms, which can then be easily linked to
other individual-level measures. This approach, however, mainly focuses
on structural properties at the global level rather than sequential
patterns for individual sequences.

\subsection{\emph{\textbf{Process mining models}}}

Process mining models are designed to study processual phenomena using
event data, particularly in the fields of business process management
and information systems (\cite{VanDerAalst1998,VanDerAalst2005,Pentland2020,Berti2023,Franzoi2023}).
The primary goal is to distill a structured description of processes
based on real-world event logs (\cite{VanDerAalst2005}). It is
believed to be particularly valuable for exploring questions regarding
temporal features in the processual phenomena, different ways a process
can be performed, and the systematic drift of processes (\cite{Franzoi2023}).

In the proposed framework, process mining could provide a novel way to
discover and monitor activities recorded in digital trace data. For
instance, a previous study (\cite{Sarirah2021}) investigated user
navigation on a Malaysian news website based on four weeks of event
logs. They found that users tended to access the website via article
pages instead of the front page. Moreover, sports and entertainment are
the sections that are most accessed in the user navigation process.
However, process models generally require clearly defined process
elements and can become considerably complex (which do not provide many
actionable insights) (\cite{Imran2022,Franzoi2023}).

\subsection{\emph{\textbf{Language-based models}}}

Natural Language Processing (NLP) techniques have revolutionized the way
sequences of words are transformed into meaningful measures (\cite{Evans2016}). In computational linguistics, researchers have
extensively used language models to study the syntactic and semantic
structures in texts, such as Word2Vec and Transformer (\cite{Vaswani2017,Mikolov2013}). These models typically assume a structuralist
perspective that the meaning of words arises from contexts in the text
(\cite{Rodriguez2023}). Recently, language-based models have evolved
rapidly and have been proven successful at capturing structure in
sequences beyond human language, including music, networks, and life and
research trajectories (\cite{Grover2016,Huang2018,Murray2023,Fan2024,Savcisens2024}).

Similar to how word embedding spaces have made breakthroughs in our
understanding of human language, these models can help us develop a new
understanding of user-sequences. Specifically, these models allow us to
avoid hand-crafted features and produce a compact, dense, and continuous
vector representation of event logs, which compresses contextual
information within user-sequences. These vector-space representations
could potentially provide a robust framework for quantitative
explorations of sequence-level patterns, similar to other application
cases (\cite{Fan2024,Savcisens2024}). In addition, the
individual-level representation can be linked to information gathered
through other approaches, such as surveys and experiments. For instance,
by incorporating these complex patterns from individual user-sequences
into modeling media effects while controlling for other covariates, we
can potentially make causal claims grounded in a robust understanding of
the structure and relationships among individual activities on digital
platforms.

Despite the vast possibilities, it is important to note that such
language-based models are powered by neural embedding techniques, which
are black-box in nature and can be computationally intensive. In
addition, these models require a large amount of high-quality data
(\cite{Savcisens2024}).

\section{\textbf{Data}}

To illustrate the proposed framework, we applied the aforementioned
approaches to a massive dataset collected through data donation. Data
donation utilizes users\textquotesingle{} right to platform data (based
on EU GDPR), which requests users to download data stored on platforms
about them and donate data to researchers (\cite{Ohme2024}). As part
of a large 2-wave panel study focusing on German platform users, we
recruited a representative sample of 2,457 participants, including an
oversampling of young adults (18-27) by 16.37\% via the panel company
Bilendi. After extensive informed consent, which explained the purpose
of data collection, participants who agreed to donate their data were
requested to follow the detailed instructions and download copies of
their digital traces. Notably, participants had the option to delete
certain personal records and could choose to donate data from one or
more of the four studied social media platforms---Facebook, YouTube,
TikTok, and Instagram.

We obtained 521 data donations from 347 unique users, containing
5,653,820 timestamped digital traces. Given the retrospective nature of
the data donation approach, the starting dates of platform data vary
across participants, but all donations ended on September 4, 2024, at
the latest. In the present article, we focus on a subset of the dataset
spanning from May 1, 2024, to August 1, 2024. This sample includes
1,262,775 records from 309 unique users over three months and across
four platforms, with 208 users donating data from one of the platforms,
67 from two platforms, 26 from three platforms, and eight users
providing data from all four platforms. On the platform level, our
sample includes data from 117 Facebook users, 140 Instagram users, 54
TikTok users, and 141 YouTube users. For a detailed overview of the
dataset, see the SI Dataset.

Drawing on this original set of digital trace data, we represent the
evolving user activities as chronologically ordered user-sequences
(Figure 1). We observe that the length distribution of these sequences
is right skewed, with a long tail towards a large volume of activities
in the donated data download packages. Specifically, we find that over
half of the users donated fewer than 1,000 records, and one donated
83,372. The average number of records per user is around 4,000, driven
by the extreme values. The relatively large number of observations for
many users and the high time resolution allow us to apply multiple
approaches to explore sequential patterns in the hyper-longitudinal
data.

\section{\textbf{Application}}

Having described the dataset, we will next focus on applying the six
approaches to our sample. The focus of our case study is to give a
straightforward example of the methods rather than addressing specific
research questions. Hence, rather than zooming in on one particular
behavior (e.g., activity switch), we are zooming out to consider how
different methods can potentially uncover patterns at different levels
for diverse research purposes (e.g., whole sequences, latent states,
durations, and subsequences). As shown in the previous work, defining
the \emph{lexicon} of events---the set of event types---is a critical
step in sequence-based approaches (\cite{Mahringer2021}). Below,
we study some basic attributes of user-sequences from the raw DTD, such
as platform and activity type (likes, watches, and searches).
Importantly, we note that the multi-platform data points are not
directly comparable (see the 5. Data alignment, Practical challenges
section). In the current paper, we study user-sequences with a simple
lexicon of platform or activity types from raw data to showcase these
approaches. Further data processing and alignment work is required for
more rigorous multi-platform research, such as defining what constitutes
\emph{active engagement} on different platforms of diverse affordances.
We provide a detailed overview of the data points and further reflection
in the SI Dataset. Additionally, these approaches can be used to examine
other attributes like content-based features (through data linkage and
augmentation). Since visual representation has been essential for
sequence-based approaches, we used a consistent color scheme across all
figures. To facilitate future research, we provide all the code we used
to wrangle, analyze, and visualize data in an online repository.

\subsection{\emph{\textbf{Sequence analysis}}}

We used sequence analysis to compare the user-sequences in terms of
platform and activity types (e.g., "Facebook\_searches" and
"Instragram\_likes") and examine the modality of users transitioning
among different activity types. First, we visualized and analyzed these
sequences. Practically, we found that the skewness in the length
distribution makes it challenging to align the sequences. To mitigate
this issue, we focused on 50\% (151) of the users falling within the
25\textsuperscript{th} to 75\textsuperscript{th} percentile of sequence
lengths.

As shown in Figure 2a, we plotted the subset of user-sequences sorted by
the starting event. It is evident from this graph that this subset still
varies considerably in length, ranging from 229 to 2,802 elements. The
graph also indicates that most sequences mainly consist of activities
from a single, dominant platform, while few users engage evenly across
platforms. This reflects the bias in the data collection, as most users
donated data only from one platform.

Next, we applied the optimal matching approach (\cite{Abbott1995}) to compare
these sequences. After computing the distance matrix, we used
hierarchical clustering to classify users. Here, we finetuned the
critical hyperparameter, i.e., the number of clusters. During this
process, we found a clear trade-off in the sense that smaller numbers
tended to group users with diverse activities together, while larger
numbers tended to generate an excessive number of clusters, with some
containing only one or two sequences. After finetuning, we selected 20
as the optimal number of clusters (see Figure 2b and c for two
examples). We observe that although users generally engaged in similar
platform activities within these clusters, lower-level clusters are
extremely small in size. Therefore, the substantial variation in
sequence lengths posed challenges to the traditional sequence analysis,
especially given the fact that another half of the users have not been
studied.

As the holistic approach considers differences between whole sequences,
not within sequences, we further analyzed the most frequent subsequences
to examine users\textquotesingle{} behavior motifs in the complete set
of 308 users. In other words, the subsequence analysis aims to describe
the large number of activities in user-sequences using a small number of
patterns (i.e., inductive pattern mining). Notably, here, we focused on
a subset of important activities, including public reactions (e.g.,
likes, comments, shares), private reactions (saves), views, and
searches.

To highlight the \emph{transitional} patterns in these high-intensity
sequences, we employed a sliding window approach outlined in a previous
study (\cite{Keegan2016}), which collapses the identical consecutive
activities that occurred within 10 minutes into a single activity
session. For example, \emph{User1: (Instagram\_likes,
08:01:23)--(Instagram\_likes, 08:07:26)--(Instagram\_likes, 08:16:01)},
this activity sequence will be collapsed into an activity session:
\emph{User1(Instagram\_likes, {[}08:01:23, 08:16:01{]})}. After merging
the identical consecutive activities within the 10-minute interval, we
significantly reduced the volume of the data. Next, we explored the
behavioral motifs at multiple levels, including bigrams, trigrams, and
four-grams. As shown in Table 1, we find that the most frequent and
statistically significant consecutive activities are identical
activities. Interestingly, transitions between certain activities on the
same platform are also common motifs, such as liking to sharing on
Instagram and watching videos to favoriting content on TikTok.

Given the diverse nature of activities on different platforms (with the
YouTube video being much longer than TikTok videos), we also implemented
different sliding windows of 20 and 30 minutes, respectively. We found
that this yielded dramatically different results---TikTok activities
became the most common motifs for their high baseline frequency in the
dataset (see Figure S1). Therefore, cautions need to be exercised when
preprocessing the data and comparing activities across different
platforms.

In summary, these observations suggest that sequence analysis, while
being valuable and straightforward, can be relatively limited when
analyzing user-sequences of digital traces across different lengths and
platforms. In particular, the holistic approach mainly considers the
differences among whole sequences while risking overlooking the
transitional patterns within sequences. By contrast, the subsequence
approach can identify patterns within sequences---motifs of a limited
number of activities---while being limited in comparing these patterns.

\subsection{\emph{\textbf{Event history analysis}}}

We employed event history analysis to study the duration time of
activities within user-sequences. Our analysis focused specifically on
how users switch platforms for those who donated data from multiple
platforms. The key \emph{event} in our event history analysis is the
platform switch, akin to life and death events in traditional survival
analysis.

We analyzed the duration of each record whose key events (i.e., the
platform switch) have been observed or unobserved. In this analysis, we
used the session-level data that merged the consecutive activities on
the same platform within the 10-minute interval.

We first used the Kaplan-Meier estimator (\cite{Kaplan1958}) to
estimate the survival function. Figure 3 plots the survival curve for
each platform. We observed that most activity sessions ended within 2
hours. When comparing the duration of activities across platforms, we
find that Facebook and Instagram activities start with a quicker
decline, whereas YouTube and TikTok curves have a slower decline that
keeps users engaged on the platform. YouTube and TikTok estimates have
larger confidence intervals, indicating more uncertainty in the
estimation (as we have a small number of events). In addition, some
curves stopped above zero. This likely reflects that the data is
right-censored, in which we did not observe the event in the data.

Finally, we used the Cox model to study the relative hazard of switching
to another platform compared to the baseline. We used dummy variables to
encode the platforms as categorical covariates in the model. Table 2
summarizes the results. Overall, we observe that compared to the
baseline (i.e., duration on YouTube), sessions on Facebook and Instagram
are significantly more likely to switch platforms, whereas TikTok
sessions are slightly less likely to switch platforms in the dataset.

In summary, event history analysis can provide a dynamic picture of how
the probability of a particular event changes over time and across
different covariates. It can handle right-censoring when an event has
not occurred during observation. However, this approach focuses
specifically on well-defined events. Moreover, it does not account for
the complex dependencies between current and previous events.

\subsection{\emph{\textbf{Hidden Markov Models}}}

We used the Hidden Markov Model (HMM) to explore internal dependencies
in the sequences. Our analysis focuses on the platform and critical
activity types for multi-platform user-sequences.

We find that the trained model struggled to converge when dealing with
the entire length of these sequences. Therefore, we focused on the
daily-level patterns. To further mitigate the length issue, we only
analyzed 1,898 (65\%) daily sequences falling within the 25th to 90th
percentile of sequence lengths.

Of interest here are the hidden states and their transitions. We
determined the critical hyperparameter, i.e., the number of hidden
states, based on the Akaike Information Criterion (AIC) and Bayesian
Information Criterion (BIC) values. We find a clear minimum of these
scores when choosing four hidden states. After determining the optimal
number of hidden states, we examined the transition and emission
probabilities. Figure 4a plots the transition probabilities between any
of the two hidden states, with darker shades indicating higher
probabilities. We find that every state has the highest probability of
remaining in its current state.

Next, we examined the emission probabilities (Figure 4b), which show the
emission probabilities of each activity type concerning a given state.
It is apparent from this graph that each hidden state generally
corresponds to the activities on one platform---TikTok for state 0,
Instagram for state 1, YouTube for state 2, and Facebook for state 3.

Overall, we observe that HMM can effectively capture platform-based
differences in the co-occurrences of activities as hidden states. Future
studies should go beyond our focus on basic attributes to include more
refined features, such as content details. We believe this could provide
deeper insights into the underlying mechanisms behind the activity
transitions in user-sequences.

\subsection{\emph{\textbf{Network analysis}}}

Next, we examined the network structure in user-sequences. We
constructed weighted and directed networks of platforms and activities
based on the complete set of sequences and activities, where two
elements are connected if they appear sequentially. The link strengths
were determined based on the frequency of transitions between the two
elements within user-sequences.

We first examined the platform-level network (Figure 5a). All nodes are
connected, signaling the existence of every platform switch in our
dataset. A closer inspection of the network shows that the strengths of
the connections between any of the platform pairs do not have a strong
directional bias. In other words, we found that no one-way switch was
much stronger than their reciprocal counterparts.

Next, we examined the network of activities, which unfolds the
within-platform behavior patterns. We find that activities across
different platforms are well-connected in the network. The strongest
links are those within the same platforms, e.g., the link between
watching and searching activities on YouTube. This aligns with the
findings from Table 1 and Figure 4a. We further computed the importance
of activities based on network metrics. Our analysis revealed that the
TikTok login history has the highest in-degree centrality. This suggests
that login on TikTok has the largest incoming attention flow from other
nodes. Regarding the out-degree centrality, the TikTok video-watching
activity has the highest outgoing flow. Moreover, TikTok login history,
Instagram likes, and YouTube watch history have the highest closeness
centrality, suggesting that these activities are the closest to every
other node in the network.

Finally, we detected communities of activities in the network. We
applied the Clauset-Newman-Moore greedy modularity maximization
algorithm (\cite{Clauset2004}) to identify the community structure
with the highest modularity on the weighted and directed network. The
modularity score is 0.34, suggesting a moderate level of division of the
network into subsets of activities (\cite{Newman2006}). Specifically, the
graph partition algorithm identified two communities. The first
community contains all kinds of activities on TikTok and YouTube
watching activity, and the second community consists of all the
remaining activities, including those on Facebook and Instagram, as well
as YouTube search. This suggests that TikTok is rather self-contained
and isolated from the other platform activities in the dataset.

Taken together, these results indicate that network analysis can uncover
structural dependencies among these activities, grounded in the relative
frequency of sequential adjacent pairs of activities in the entire
dataset of user-sequences. Moreover, it can be further extended to other
forms (e.g., dynamic network, user-level network, and bipartite
network), which can serve as an important basis for understanding user
behaviors. However, unlike other methods introduced in the paper, it
mainly focuses on the global configuration of activities within the
sequences constituting a complex system of diverse activities. This,
however, does not explicitly account for the sequentiality on the
individual level.

\subsection{\emph{\textbf{Process mining models}}}

We employed process mining tools to extract information from event logs
on a daily level. By doing that, we aimed to uncover typical
users\textquotesingle{} daily routines on these platforms. Here, we
focused on the routine of multi-platform users in terms of platforms and
activity types. Specifically, we seek to understand how those users
switch platforms using process discovery algorithms.

We first explored the processual patterns in the multi-platform
user-sequences, focusing on the subset of key activities. We observe
that on the platform level, the process model shows four platform nodes
with no clear order of switches. This corroborates the findings from the
network analysis (Figure 5a). Next, we examined the time dimension
regarding the average time users spend transitioning among four
platforms within a single day. Figure 6a shows these transition times,
with darker shades indicating longer times. Unsurprisingly, we found
that the transition time within the same platform is significantly
shorter than cross-platform transitions. For example, TikTok users
typically return to it within 4 minutes. Notably, switches from Facebook
to Instagram took, on average, 3.4 hours, whereas transitions from
Instagram to Facebook required more than 2.7 times longer (\textgreater9
hours, on average) within a day.

Next, we zoomed into the processual patterns on the level of activity
types. Here, we focused on users\textquotesingle{} digital routines. We
found that the process model was becoming extremely complex when
analyzing all paths in the sample. In order to provide a simple
illustration of the approach, we focused on the top ten daily paths. We
also added an artificial node, "platform\_switch," to highlight
switching behaviors. Figure 6b shows the business process modeling
notation (BPMN) graph. Based on this process map, we identified two main
types of user paths:

\begin{enumerate}
\def\labelenumi{\arabic{enumi})}
\item
  The TikTok path: Users begin their digital routine by watching videos
  on TikTok. Following this, they may like what they have watched, which
  could lead them to return to watching more videos. Eventually, users
  may either continue watching additional videos or end their TikTok
  session.
\item
  Facebook and Instagram paths: Users started by interacting on Facebook
  or Instagram. For instance, if they first like content on Instagram,
  they are likely to switch platforms to Facebook in the next step.
  After being on Facebook, they are likely to search and react to
  content before ending their session (as indicated by the parallel
  gateway +).
\end{enumerate}

Interestingly, watching videos on YouTube typically occurs at the end of
these user\textquotesingle s daily path. In other words, if users use
YouTube, it typically appears at the end of their daily platform
routine.

Figure 6c is a directed follow graph that provides further details on
the process model. By concentrating on the strongest path (thick lines),
we observe that from the starting point, most paths (14) moved to like
some content on Instagram, and the majority of the liking activities (14
of 16) on Instagram led to platform switch behavior. Switching platforms
(28) would most likely lead to searching on Facebook (10). This aligns
with the identified Instagram path in the BPMN graph.

Overall, the process mining approach provides interesting insights into
procedural phenomena in user-sequences. However, the application of this
method requires considerable attention to data preprocessing, and it can
be challenging to identify the appropriate model for complex data,
particularly when balancing critical model properties such as
simplicity, fitness, and generalization (\cite{VanDerAalst2012,Imran2022}).

\subsection{\emph{\textbf{Language-based models}}}

We generated a 100-dimensional vector space representation of the event
logs in the complete set of user-sequences. In particular, we created a
synthetic language, which combines the platform name, activity type, and
limited content information (e.g., search terms and URLs) to form a
vocabulary. For example, a synthetic word is "Facebook\_searches\_Events
- Dachau / Umgebung." As shown in Figure 7, the learned 100-\emph{d}
representations of event logs are projected onto a two-dimensional space
using \emph{t}-distributed stochastic neighbor embedding (t-SNE) (\cite{VanDerMaaten2008}). It is clear from the graph that the overall
structure is organized based on the platforms, as activities on the same
platform tend to group together as distinct clusters. Digging deeper
into some regions of the graph, we find that the model learned
non-trivial associations between similar activities. For instance, in
the highlighted blue region, we observe a compact local structure,
suggesting that activities are closely related to each other. We
discovered that these Facebook searches are about nearby towns in
Germany, such as Olching, Fürstenfeldbruck, Dachau, and Maisach,
suggesting that geographical similarity in search terms is captured as a
form of spatial proximity in the vector space. In another highlighted
region of grey points, we find that these TikTok videos generally focus
on entertainment and pop culture, including Japanese Day and Korean
animation, as well as advertisements for the gaming mouse, Samsung
watch, and Disney Plus. This also reflects certain levels of
personalization and algorithmic curation on TikTok.

In this simple application case, we treated each synthetic word as a
token with no contextual understanding of the content. Future research
should incorporate additional features (e.g., content types) and explore
the application of more complex model architectures (e.g., with
attention mechanisms). Even so, as illustrated in this case study, a
simple neural embedding model enables us to effectively encode detailed
relationships among activities from user-sequences. Based on the
obtained vector-space representation, we could reconstruct individual
user-sequences, composed of consecutive synthetic words (Figure 7). This
could potentially provide a powerful framework for further quantitative
examination of sequential patterns. For example, we can group users
based on similar movement patterns in the vector space. Alternatively,
we can characterize each user-sequence in terms of their movements
within the space using methods from mobility analysis---such as entropy
(\cite{Shannon1948}), which captures the predictability of activity over
time, or the radius of gyration (\cite{Gonzalez2008}), which
characterizes how far a user moves from the central point.

Additionally, we can finetune each sequence and compress it into a
person-level summary in the form of a single vector for specific
prediction tasks (\cite{Savcisens2024}). In particular, this approach
allows us to effectively encapsulate the most relevant features in
user-sequences and produces a consolidated representation for predicting
specific outcomes, such as those measured in the survey.

However, it is important to note that these models require massive,
high-quality datasets for training. Only with rich data, these complex
models could uncover patterns encoded within user-sequences and
transform the data into meaningful and generalizable individual-level
representations.

\section{\textbf{Discussion and conclusion}}

The possibility of gathering digital trace data can transform the
empirical basis for communication research. Given that digital platforms
are the most frequently used communication spaces for most users
nowadays, it is crucial to not only develop methods to capture DTD but
also find new ways to analyze them. This study proposes a new framework
for analyzing hyper-longitudinal digital trace data as time-evolving
user-sequences. By applying six approaches to these sequences, this
study examines the interplay between analytical challenges posed by the
data and the possibilities brought by different approaches. Table 3
lists a detailed assessment of how well each approach can address these
challenges.

Overall, there are three important learnings from this study: Digital
trace data, so far, have only been collected from subsamples, thereby
missing insights into representative datasets, as the data collection
methods themselves often produce sample biases (\cite{Wedel2024}). Our
study shows that many approaches (e.g., sequence analysis, HMM, and
process mining) require significant processing, which further reduces
the amount of data being studied, either because of too-long timeframes,
too-long sequences, or missing data. This underlines the challenge of
arriving at generalizable statements based on DTD.

Second, we see that the raw DTD we used (which has yet to be linked or
augmented with content or other metadata) provides highly granular
insights, especially for cross-platform research, which is a blank spot
in many fields of digital communication research. Specifically, many
approaches (e.g., subsequence analysis, network analysis, and process
mining) can predict the switch from specific events within a platform
(from a like to a share) or from exposure on one platform to another
platform. They can also model a typical daily routine of platform users,
which is unprecedented in platform research and has mostly been covered
by interview studies. Our study presents only the first step, as we
mainly focus on some basic attributes of raw DTD. In the next step, we
believe further data alignment and processing work and augmenting raw
DTD with content information, such as classified multimodal content
data, will make the modeling of user-sequences within and across
platforms more insightful. This moves from raw activity data to the more
granular and meaningful constructs.

Third, the language-based model stood out for its ability to capture
sequence-level patterns and deal with most of the challenges. This
approach is especially helpful \emph{if} applied to large-scale,
high-quality data. Moreover, language-based models can help make causal
claims in communication research because they can encapsulate a
person-level summary that encodes essential sequence-level information
\emph{if} we link large-amount, multimodal, high-resolution trace data
to important outcome variables measured through other approaches
(similar to how \cite{Savcisens2024} study mortality by applying
transformers to sequential life-event data). We believe that the biggest
steps are necessary with this type of method for its great potential.
However, it draws our attention to one big \emph{IF}. High-quality
datasets that detail how individual sequences evolve in the space of
diverse platform activities (e.g., information about a comment made by
the user is coupled by a topic switch or the time of the day) are
necessary for such research. As they do not exist at the moment, this
calls for more global and collaborative data gathering and training
efforts for the study of DTD to arrive at training datasets that can
start a new era for platform research.

This study has a number of limitations. Although we use one of the most
comprehensive user-centric digital trace datasets, we have to
acknowledge that it can only make a statement based on a subsample of
the German quota sample. Specifically, we lack users who have shared
their data from three or four platforms. This unequal representation of
the number of platforms in DDPs makes applying some analytical
approaches more challenging. Moreover, the set of approaches presented
in this study is not exhaustive. Future studies should go beyond our
focus to use other sequence mining methods. In addition, our study only
employs raw DTD, which underestimates the challenges of data alignment
and further processing, both in terms of the capacities and the
challenges arising from them. Hence, this study provides a more
optimistic view of how these approaches can be employed to analyze
user-sequences for addressing actual communication research questions.

Our study nonetheless provides important insights for scholars striving
to study longitudinal digital trace data. It can guide the selection of
approaches, along with associated challenges. Without giving definite
answers, we hope this piece can spark discussions within the field about
the optimal ways to analyze hyper-longitudinal digital trace data---the
race has just begun, and a long way still lies ahead of platform
researchers.

\section{\textbf{Acknowledgments}}

This research was funded by the Digital News Dynamics (DND) research
group at Weizenbaum Institute (WI), Berlin. The WI is funded by
Bundesministerium für Bildung und Forschung (BMBF), Germany, and Land
Berlin, Germany.

\section{\textbf{Data availability}}

The dataset cannot be provided due to its sensitive nature. The code has
been published in an online repository. We provide anonymous code and
material to reviewers through the Open Science Framework (OSF):
\url{https://osf.io/y2nva/?view_only=f4bd3eb8ba48485d990abbe3112a05e5}.
Reviewers will find them in the files section of the anonymous OSF
project.

\section{Figures and tables}

\begin{figure}[H]  
    \centering
    \includegraphics[width=\textwidth]{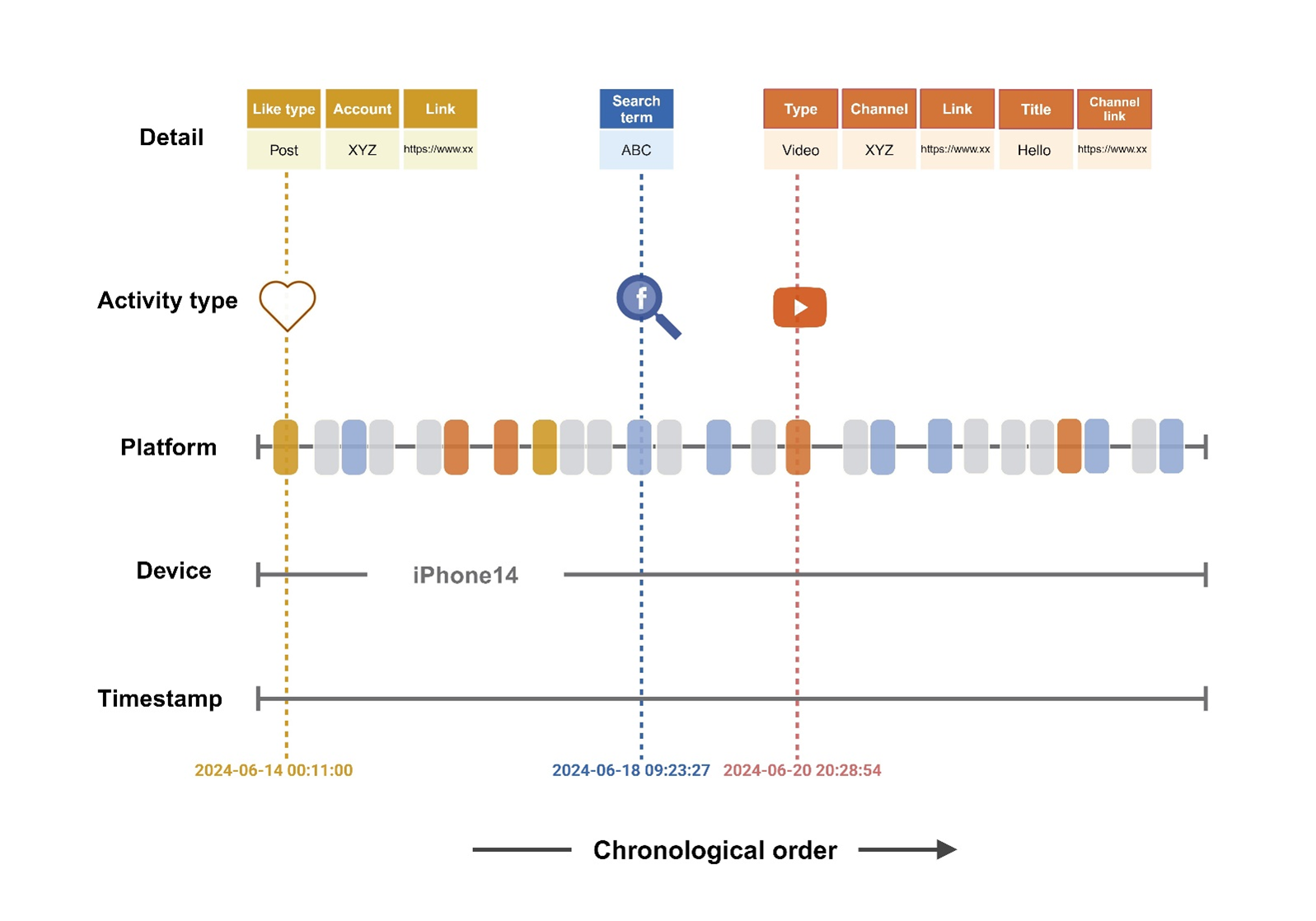}
    \caption{\textbf{An example of individual-level data representation.} Here, we organize the multi-platform digital trace data into a chronologically ordered user-sequence. Each block represents an activity, and the color of the block indicates the platform. Each activity is associated with a timestamp and other information. \label{fig:1}}
    
\end{figure}
\begin{figure}[H]  
    \centering
    \includegraphics[width=0.8\textwidth]{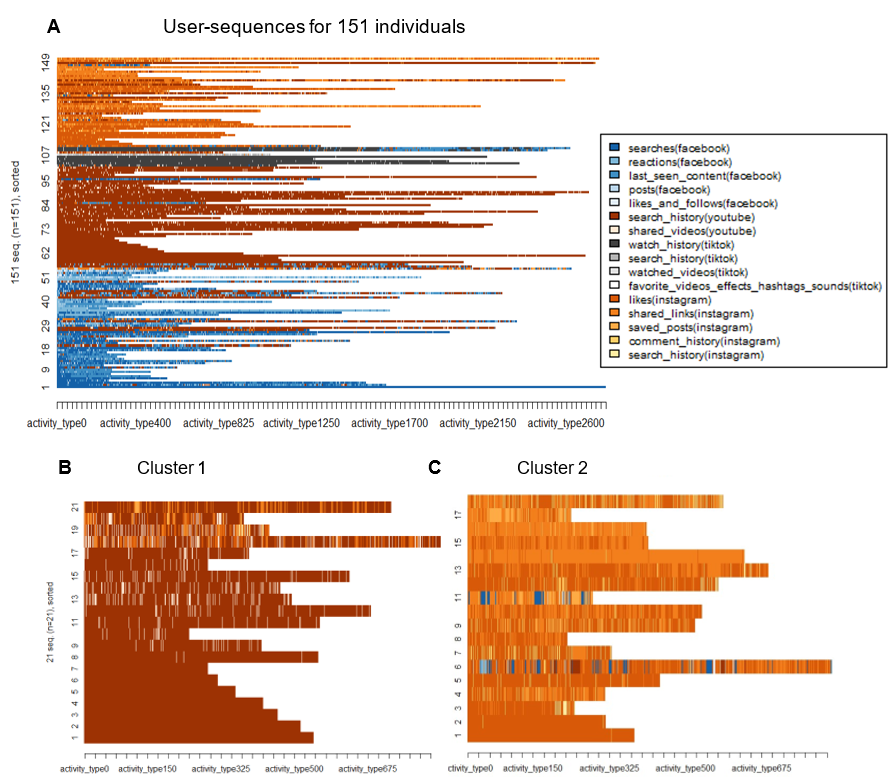}
    \caption{\textbf{Sequence index plot of platform and activity types.} The plots use colored stacked bars to show how users move between different activities over time. Each horizontal bar represents a user, and the segments' colors indicate the platform and activity type.  \label{fig:2} }
 
\end{figure}

\begin{figure}[H]  
    \centering
    \includegraphics[width=0.8\textwidth]{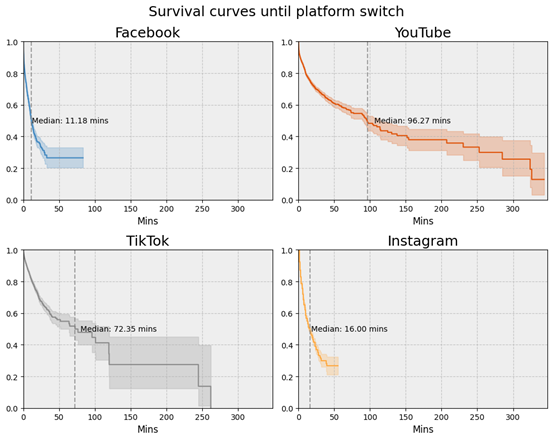}
    \caption{\textbf{Survival curves for users remain on the same platform.} For each subplot, the x-axis represents the duration time t, and the y-axis represents the probability that a user is still on this platform after t minutes. \label{fig:3} }
    
\end{figure}

\begin{figure}[H]  
    \centering
    \includegraphics[width=0.7\textwidth]{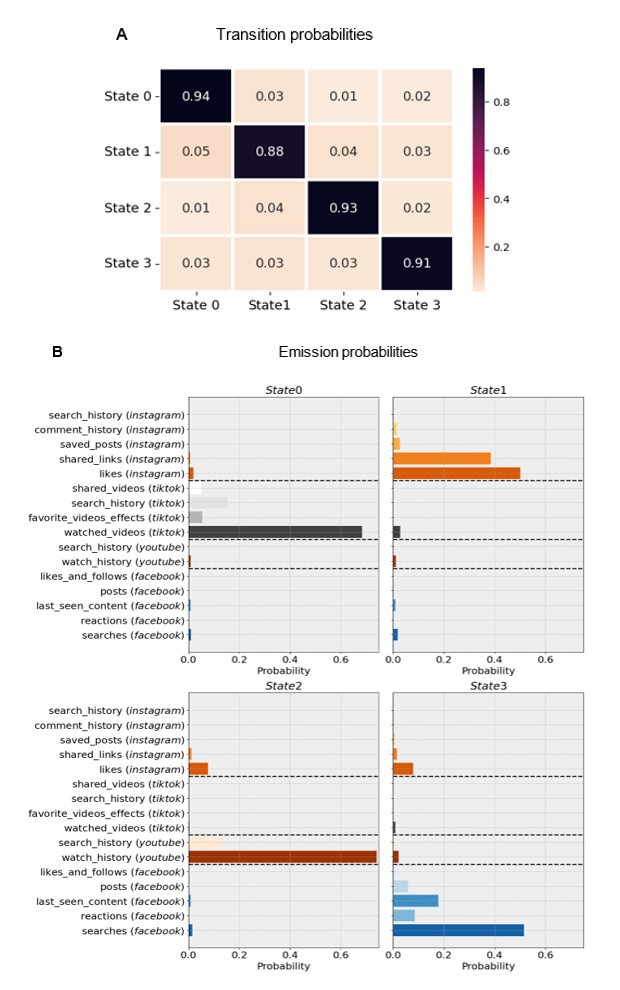}
    \caption{\textbf{Transition/emission probabilities among the four learned hidden states.} (a) the heatmap shows the probability of transitioning from one hidden state to another, with the darker colors indicating a higher probability. (b) the bar plots show each hidden state with emission probabilities for each activity type. Here, each bar represents the probability of observing a specific activity given the hidden state. \label{fig:4} }
    
\end{figure}
\begin{figure}[H]  
    \centering
    \includegraphics[width=0.8\textwidth]{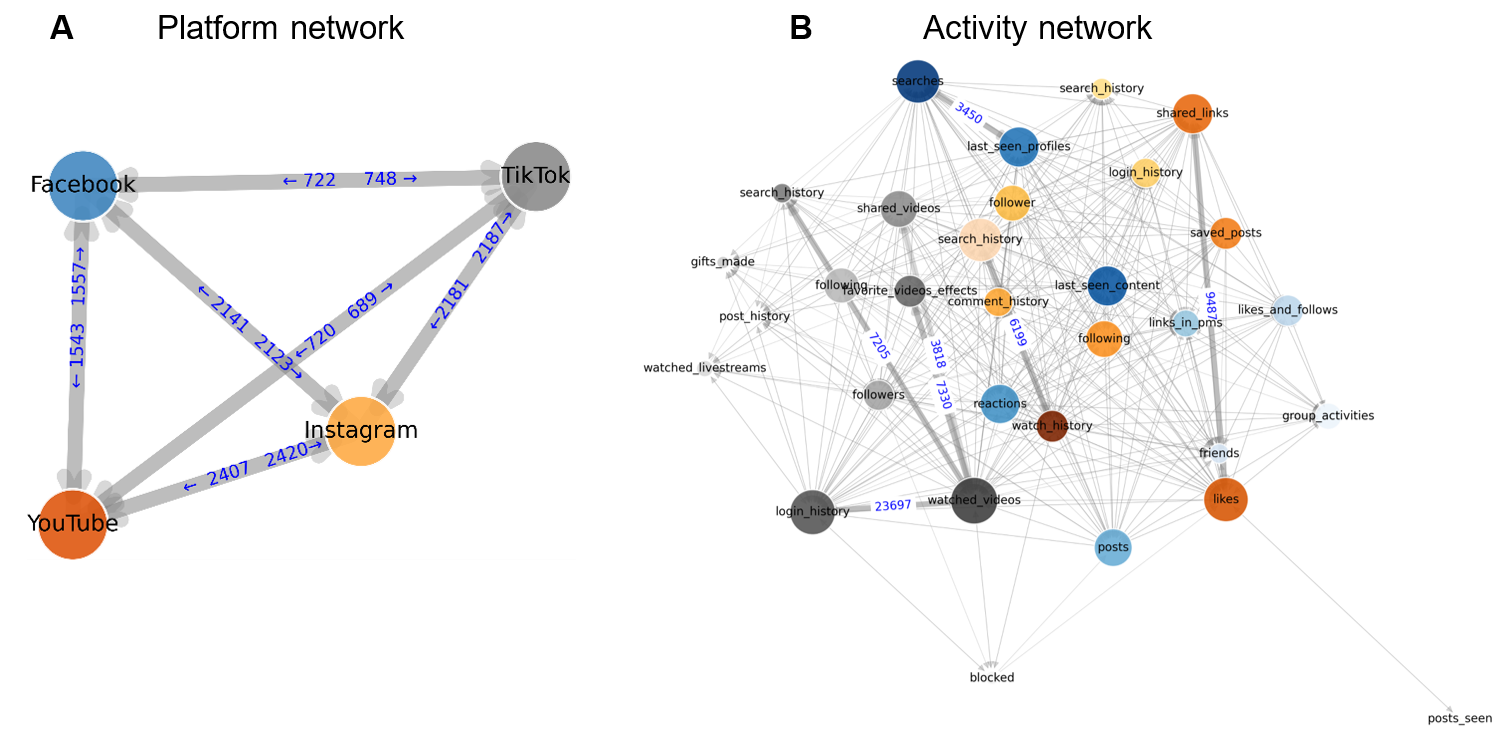}
    \caption{\textbf{Networks of platforms/activities.} Here, each node represents a platform or an activity. The links connect two nodes when they appear sequentially in user-sequences. The nodes are colored by platforms and sized by degree centralities.   \label{fig:5}}
  
\end{figure}

\begin{figure}[H]  
    \centering
    \includegraphics[width=\textwidth]{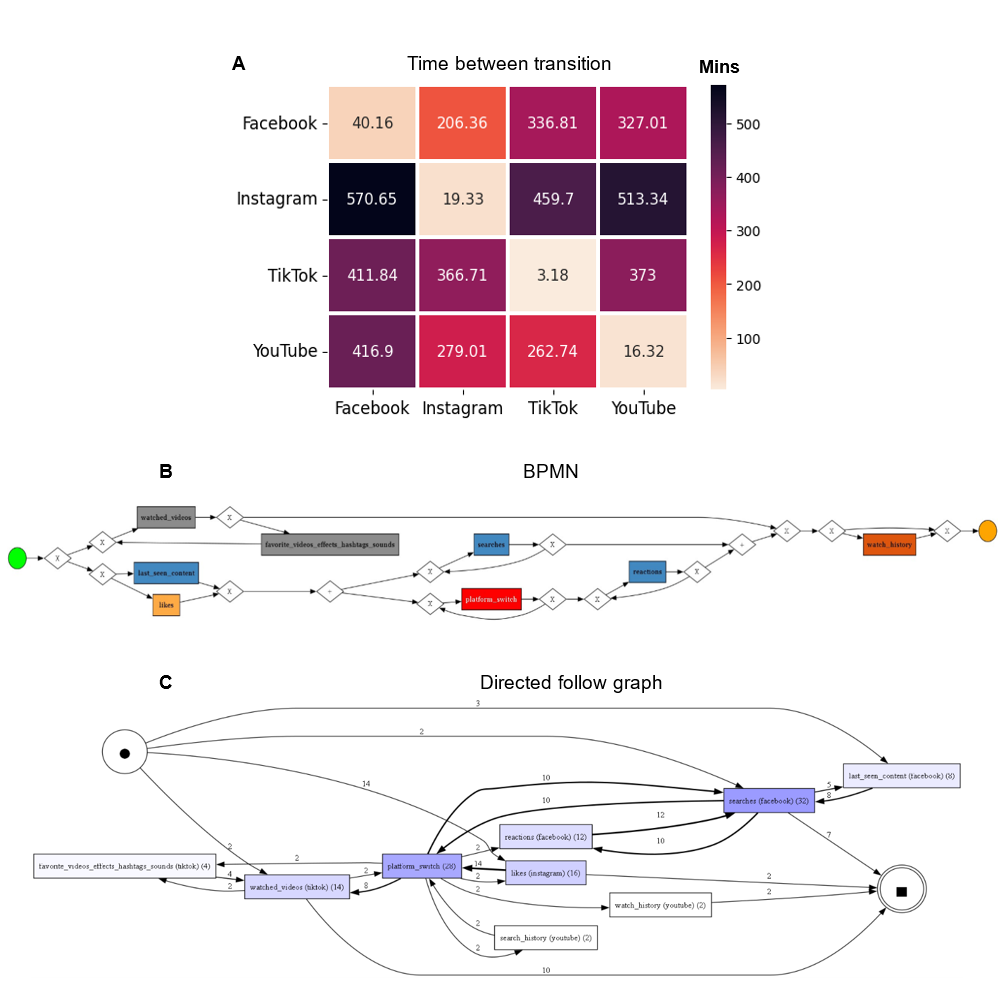}
    \caption{\textbf{Process mining of daily event logs.} (A) the time between transitions across platforms. (B) the BPMN graph. Here, the path starts from the green node and proceeds to an exclusive gateway, X, to either the TikTok path or the Facebook and Instagram path, converges to either watch YouTube videos or bypass it, and finally ends at the yellow node. In particular, we added an artificial node to highlight the platform switch. (C) the directed follow graph. Here, the nodes represent activities, and directed links represent the sequence of activities in the process.   \label{fig:6} }
  
\end{figure}

\begin{figure}[H]  
    \centering
    \includegraphics[width=0.8\textwidth]{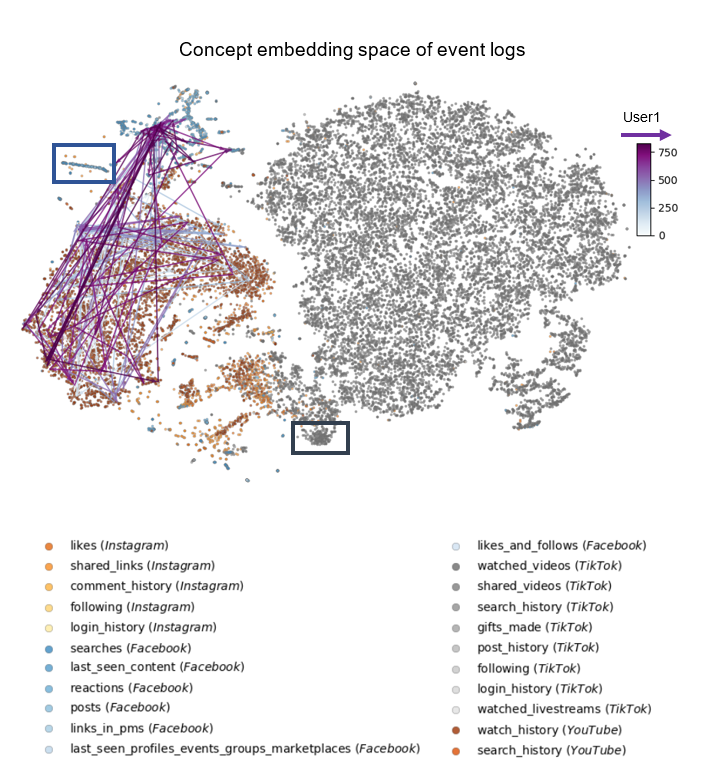}
    \caption{\textbf{Two-dimensional projection of the event logs using t-SNE.} Each point represents an event log, which includes platform name, activity type, and content information. Points are colored based on the activity type. We have highlighted two regions as examples. The blue region represents searches on Facebook, consisting of search queries about nearby areas in southern Germany, including "Du kommst aus Olching, wenn...," Veranstaltungen in Fürstenfeldbruck und Umgebung," and "Events - Dachau / Umgebung." The grey region represents TikTok video consumption, which generally focuses on entertainment and pop culture. We further present a user sequence within the learned vector space, which comprises 830 unique activities from an individual across two platforms (Facebook and Instagram) over three months.  \label{fig:7} }
\end{figure}
\subsection{Tables}
\begin{figure}[H]  
    \textbf{Table 1. Most frequent subsequences. All subsequences in the table appear significantly above chance at \newline \emph{p} < .0001 with Bonferroni correction. }
    \centering
    \includegraphics[width=\textwidth]{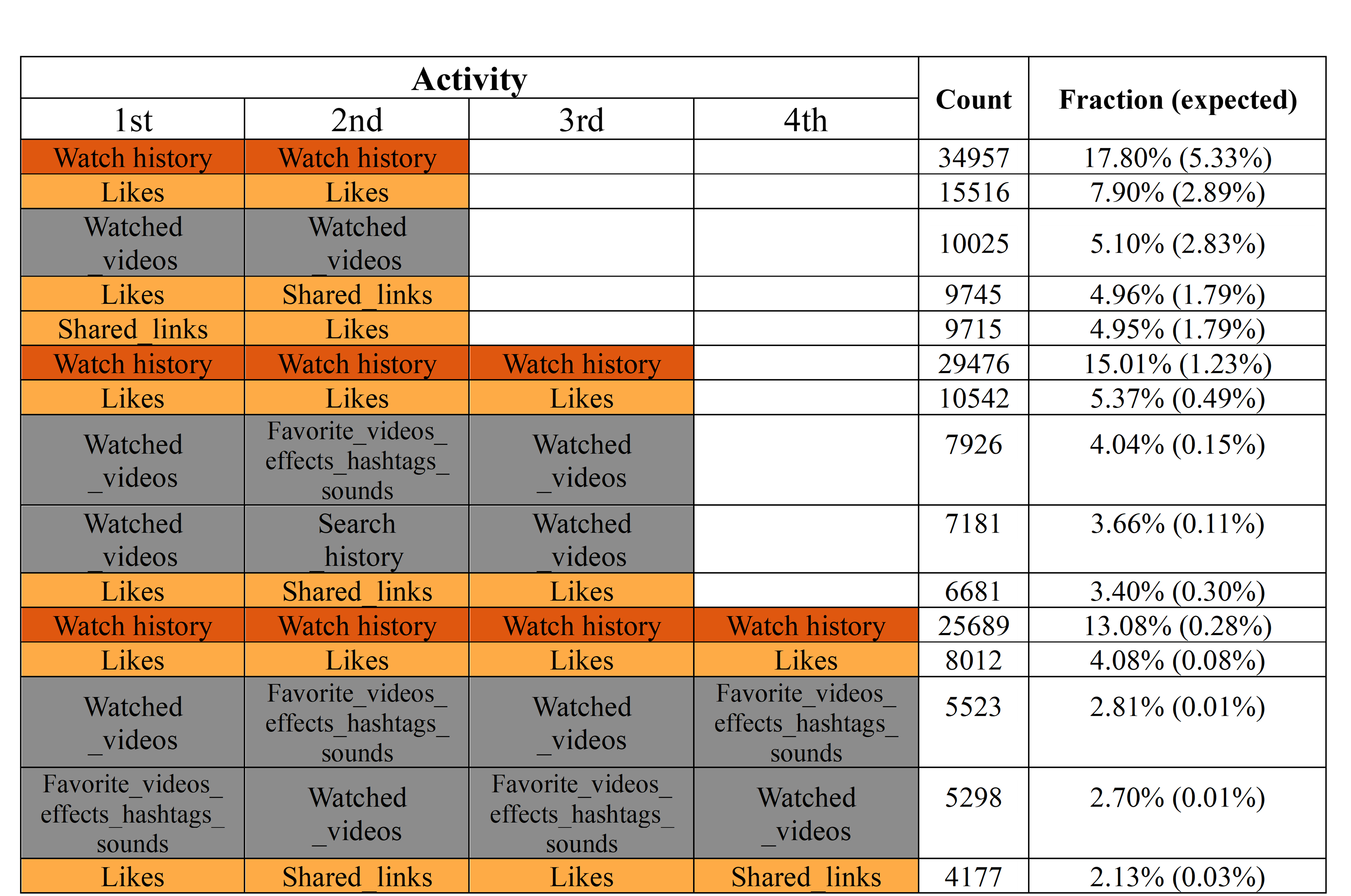}

    \label{fig:table1}
\end{figure}

\begin{table}[htbp]
\centering
\textbf{Table 2. The Cox hazard model estimation }

\label{tab:my_label}
\begin{tabularx}{0.4\textwidth}{l c c c}  
\toprule
\textbf{} & \textbf{co ef} & \textbf{se(coef)} & \textbf{\emph{p}} \\
\midrule
Facebook   & 1.06  & 0.03 & <.001 \\
TikTok     & -0.13 & 0.03 & <.001 \\
Instagram  & 0.72  & 0.03 & <.001 \\
\bottomrule
\end{tabularx}
\end{table}

\begin{table}[htbp]
\centering
\textbf{Table 3. Approaches vs.\ challenges}           
\label{tab:approaches_challenges}

\begin{tabularx}{1.05\textwidth}{%
  >{\raggedright\arraybackslash}p{3.5cm}   
  *{6}{>{\raggedright\arraybackslash}X}     
}
\toprule
\textbf{Challenge} &
\textbf{Sequence analysis} &
\textbf{Event history analysis} &
\textbf{Hidden Markov Models} &
\textbf{Network analysis} &
\textbf{Process mining} &
\textbf{Language-based models} \\
\midrule
\textbf{Volume} & Limited & Well-suited & Can be challenging with longer sequences & Well-suited & Well-suited & Well-suited \\
\textbf{Complexity} & Limited & Limited & Can handle sparsity. But limited in handling high-dimensional data & Well-suited & Limited in handling high dimensional data & Well-suited \\
\textbf{Incompleteness and skewness} & Limited & Can handle right-censored data & Well-suited & Well-suited & Limited & Well-suited \\
\textbf{Time window} & Less flexible & Well-suited & Well-suited & Well-suited & Well-suited & Well-suited \\
\textbf{Data alignment (multi-platform and survey data)} & Requires preprocessing; patterns in sequences can be clustered and linked to survey data & Requires preprocessing; can be linked to survey data & Requires preprocessing; can be linked to survey data & Requires preprocessing; limited & Requires preprocessing; can be linked to survey data & Well-suited \\
\textbf{Data volatility} & Limited & Limited & Limited & Limited & Limited & Limited \\
\textbf{Additional limitations} & Sensitive to noise, the trade-off between the pattern length and the generalizability of patterns (whole sequence vs.\ subsequence) & Focuses on specific events with typically binary (categorical) outcomes, DTD did not record the duration of events & Assumes the Markov property (one-step dependency); computational complexity & Does not account for sequential patterns explicitly & Limited if the data lacks a clear process structure; increasing complexity with volume & Requires a large amount of high-quality data; requires validation; black-box \\
\bottomrule
\end{tabularx}
\end{table}

\clearpage
\printbibliography

\end{document}